\documentclass[sigconf, screen]{acmart}  %
\AtBeginDocument{%
  }

\copyrightyear{2024}
\acmYear{2024}
\setcopyright{cc}  %
\setcctype[4.0]{by}  %
\acmConference[IUI '24]{29th International Conference on Intelligent User Interfaces}{March 18--21, 2024}{Greenville, SC, USA}
\acmBooktitle{29th International Conference on Intelligent User Interfaces (IUI '24), March 18--21, 2024, Greenville, SC, USA}
\acmDOI{10.1145/3640543.3645142}
\acmISBN{979-8-4007-0508-3/24/03}

\begin{document}

\title{iScore: Visual Analytics for Interpreting How Language Models Automatically Score Summaries}

\author{Adam Coscia}
\orcid{0000-0002-0429-9295}
\affiliation{%
  \institution{Georgia Institute of Technology}
  \streetaddress{North Avenue}
  \city{Atlanta}
  \state{Georgia}
  \country{USA}
  \postcode{30332}
}%
\email{acoscia6@gatech.edu}

\author{Langdon Holmes}
\email{langdon.holmes@vanderbilt.edu}
\orcid{0000-0003-4338-4609}
\author{Wesley Morris}
\email{wesley.g.morris@vanderbilt.edu}
\orcid{0000-0001-6316-6479}
\affiliation{%
  \institution{Vanderbilt University}
  \streetaddress{2201 West End Ave}
  \city{Nashville}
  \state{Tennessee}
  \country{USA}
  \postcode{37235}
}%

\author{Joon Suh Choi}
\orcid{0000-0002-7732-0366}
\affiliation{%
  \institution{Georgia State University}
  \streetaddress{33 Gilmer Street SE}
  \city{Atlanta}
  \state{Georgia}
  \country{USA}
  \postcode{30303}
}%
\email{jchoi92@gsu.edu}

\author{Scott Crossley}
\orcid{0000-0002-5148-0273}
\affiliation{%
  \institution{Vanderbilt University}
  \streetaddress{2201 West End Ave}
  \city{Nashville}
  \state{Tennessee}
  \country{USA}
  \postcode{37235}
}%
\email{scott.crossley@vanderbilt.edu}

\author{Alex Endert}
\orcid{0000-0002-6914-610X}
\affiliation{%
  \institution{Georgia Institute of Technology}
  \streetaddress{North Avenue}
  \city{Atlanta}
  \state{Georgia}
  \country{USA}
  \postcode{30332}
}%
\email{endert@gatech.edu}

\renewcommand{\shortauthors}{Coscia, et al.}

\begin{abstract}
  The recent explosion in popularity of large language models (LLMs) has inspired learning engineers to incorporate them into adaptive educational tools that automatically score summary writing.
  Understanding and evaluating LLMs is vital before deploying them in critical learning environments, yet their unprecedented size and expanding number of parameters inhibits transparency and impedes trust when they underperform.
  Through a collaborative user-centered design process with several learning engineers building and deploying summary scoring LLMs, we characterized fundamental design challenges and goals around interpreting their models, including aggregating large text inputs, tracking score provenance, and scaling LLM interpretability methods.
  To address their concerns, we developed \textit{iScore}, an interactive visual analytics tool for learning engineers to upload, score, and compare multiple summaries simultaneously.
  Tightly integrated views allow users to iteratively revise the language in summaries, track changes in the resulting LLM scores, and visualize model weights at multiple levels of abstraction.
  To validate our approach, we deployed \textit{iScore} with three learning engineers over the course of a month.
  We present a case study where interacting with \textit{iScore} led a learning engineer to improve their LLM's score accuracy by three percentage points.
  Finally, we conducted qualitative interviews with the learning engineers that revealed how \textit{iScore} enabled them to understand, evaluate, and build trust in their LLMs during deployment.
\end{abstract}



\maketitle

\section{Introduction}
\label{sec:introduction}

The advent of large language models (LLMs) has catalyzed state-of-the-art research in the learning analytics community on advancing the capabilities of adaptive educational tools, namely automated scoring of summary writing \cite{botarleanu2022multitask, morris2023using}.
For example, LLMs can be used to automatically score a summary written on a larger body of text (Fig.~\ref{fig:llm_data}) in a variety of learning environments.
For data scientists in the learning analytics community, henceforth called learning engineers, it is extremely important to test LLMs on many different summaries and understand how the LLMs work.
However, using deep learning models introduces opaqueness into model evaluation \cite{Kulesza:2015:ExplanatoryDebugging, Ribeiro:2016:WhyShouldITrustYou}, making it difficult for learning engineers to close the loop of model development \cite{Wang:2021:HITLNLP}.
Interactively exploring how their LLMs score different summaries can help learning engineers understand the decisions on which the LLMs base their scores, discover unintended biases, update the LLMs to address the biases and mitigate the potential pedagogical ramifications of prematurely deploying untested LLM-powered educational technologies \cite{kasneci2023chatgpt}.

Understanding and evaluating LLMs is extremely challenging due to their unwieldy size and ever-growing number of parameters, making it difficult to identify the causes of performance issues and address them as needed \cite{Srivastava:2022:BeyondImitationGame}.
Transparency is critical for building trust in using LLMs \cite{Liao:2023:AITransparencyLLMs}, especially for learning engineers who are increasingly using LLMs as ``black boxes'' for downstream tasks, as well as during human-in-the-loop evaluation of LLM performance where quantitative benchmarks often fall short \cite{Bender:2021:StochasticParrots}.
Visual analytics tools are increasingly used for improving the transparency of LLMs and helping developers interpret LLM behavior \cite{Hohman:2019:VAinDeepLearning, Liu:2017:AnalysisofMLwithVASurvey}.
Yet little research at the nexus of machine learning (ML) and educational data has leveraged visual analytics for explaining ML-powered educational technologies \cite{Chen:2016:DropoutSeer, Mubarak:2021:VAClickstreamDLMOOC, Garcia-Zanabria:2022:SDA-Vis, Zhang:2023:VADropoutCounterfactual} and none have used LLMs, presenting an opportunity to collaboratively develop design principles in this space with domain experts.
We seek to build a visual analytics system that makes evaluating summary scoring LLMs more transparent for learning engineers, helping them understand and build trust in their LLMs before deployment.

Through a user-centered design process \cite{Sedlmair:2012:DesignStudyMethods} working directly with learning engineers who train, test, and deploy automatic summary scoring LLMs (Sect.~\ref{sec:design}), we synthesized several fundamental challenges around the aggregation, provenance, and scalability of visualizing writing and LLM scoring data together.
For learning engineers, characterizing quality in writing samples involves comparing differences in both text and LLM scores across multiple scoring dimensions and between multiple different samples, revisions of the same sample, and expert-scored ``ground truth'' samples.
Performing these tasks enables them to calibrate their trust in the LLM summary scores.
Increasing transparency in LLM scores requires probing model behavior externally by varying input parameters and internally by exploring model weights, as well as scaling these methods to large amounts of text while keeping learning engineers in the loop at different levels of detail.
To address these challenges in a single interface, a successful visualization system should help learning engineers scaffold the evaluation process around comparing revisions of multiple writing samples, scale multiple interpretability methods to work with large inputs and visually aggregate text at multiple levels of abstraction.

We present \textit{iScore} (Sect.~\ref{sec:system}), an open-source\footnote{\textit{iScore} models and code: \url{https://github.com/AdamCoscia/iScore}}, interactive visual analytics tool for learning engineers to upload, score, and compare multiple summaries of a source text simultaneously.
\textit{iScore} introduces a new workflow for comparing the language features that contribute to different LLM scores by structuring analysis across three coordinated views.
First, users upload, score and can manually revise and re-score multiple source/summary pairs simultaneously in the \textit{Assignments Panel}.
Then, users can visually track how scores change across revisions in the context of expert-scored LLM training data in the \textit{Scores Dashboard}.
Finally, users can compare model weights between words across model layers, as well as differences in scores between automatically revised summary perturbations, using two model interpretability methods in the \textit{Model Analysis View}.
Together, these views provide learning engineers with access to multiple summary comparison visualizations and several well-known LLM interpretability methods including attention attribution, input perturbation, and adversarial examples.
Combining these visualizations and methods in a single visual interface broadly enables deeper analysis of LLM behavior that was previously time-consuming and difficult to perform.

To validate our approach, we deployed \textit{iScore} with our collaborators, a learning analytics team, over the course of a month, and conducted follow-up interviews with the same engineers from the team with whom we collaboratively designed \textit{iScore} (Sect.~\ref{sec:evaluation}).
We first describe a case study where one of the learning engineers used \textit{iScore} to improve the accuracy of LLMs used in \textit{iTELL}, an intelligent textbook framework that can automatically score summaries of textbook sections written by learners such as students.
Using feedback from \textit{iScore}, the engineer improved the scoring accuracy of their LLM by three percentage points.
From the same experts' qualitative feedback, we report several findings.
\textit{iScore} improved understanding of how specific models work; e.g., removing the first sentence of a summary caused some models' scores to drop by up to $90\%$.
The model analysis visualizations helped engineers ``see'' what the LLMs paid attention to, while tracking changes in scores across revisions illustrated how adversarial examples can trick LLMs into giving incorrect scores.
The perturbation visualizations were unanimously considered the most useful feature for inspiring trust when deploying LLMs in critical learning environments.
Reflecting on the case study and findings, we then discuss implications for the design of future systems, lessons learned on building responsible and ethical AI for education, generalizing our techniques to other LLMs and finally limitations and future work (Sect.~\ref{sec:discussion}).

In summary, our paper contributes: (1) design challenges and user tasks for helping learning engineers evaluate automated summary scoring LLMs; (2) \textit{iScore}, an open-source visual analytics tool that aggregates and compares LLM data and LLM-scored writing samples; (3) a case study detailing how learning engineers deployed \textit{iScore} to improve their LLMs and usage scenarios that demonstrate the generalizability of \textit{iScore}; and (4) a qualitative evaluation with learning engineers describing how \textit{iScore} helped them understand, evaluate, and build trust in their LLMs during deployment.

\section{Related Work}
\label{sec:related_work}

\subsection{Automatically Scoring Summary Writing}
\label{sec:automatic_summary_scoring}
Summary writing is a valuable pedagogical tool to help learners build knowledge about a subject area, as well as for assessment purposes \cite{head1989examination, graham2015common, phillips2019beyond}.
A meta-analysis of $56$ experiments found that text summarization improved learning regardless of the knowledge domain \cite{graham2020effects}.
The reason for these learning increases may be that, when summarizing, learners are asked not only to retrieve information from the text but also to construct and organize their own schemata of the content area \cite{galbraith2018work, nelson2023discourse, silva2019writing}.
Despite the benefits, scaling summarization tasks in learning environments is challenging; providing feedback to learners is both difficult and time-consuming \cite{gamage2021peer}.
In response to this challenge, researchers have developed several strategies to automatically score summaries.
For example, Crossley et al. used indices of linguistic features in summaries as well as Word2vec similarity scores to develop a model that explained $53\%$ of score variance \cite{crossley2019automated}. 
Deep learning methods are particularly well-suited to this task, as Botarleanu et al. demonstrated by using fine-tuned large language models (LLMs) to explain $55\%$ of the score variance in their dataset \cite{botarleanu2022multitask}. 
Morris et al. improved on this work with LLMs by using principal components rather than raw scores to explain $66-79\%$ of score variance \cite{morris2023using}.
While the language features used to explain summary scoring mapped onto expectations about what makes a good summary (i.e., overlap with source text, text organization, and vocabulary), the language features that power the LLM summary scoring models are very difficult to interpret.

\subsection{Modeling Language With Transformers}
\label{sec:large_language_models}
We aim to help learning engineers interpret large transformer-based language models that automatically assign scores to written summaries of source text.
Language models learn to model the probability of a token occurring in a sequence (e.g., a word in a sentence).
Transformer-based language models work by encoding all input words in a sentence into numeric representations known as embeddings.
Attention mechanisms are then used to combine different word embeddings in a sentence together, creating new embeddings that are contextually informed by different parts of the sentence \cite{Vaswani:2017:AttentionIsAllYouNeed}.
Attention weights determine how different embeddings are combined, and they have been used to explore how transformers work \cite{Clark:2019:AnalyzingBERTAttention}.
Large transformer-based language models such as BERT \cite{Devlin:2019:BERT} and GPT-3 \cite{Brown:2020:GPT3} have demonstrated state-of-the-art performance on a variety of tasks \cite{Srivastava:2022:BeyondImitationGame} in part due to pre-training via self-supervised learning using large-scale unlabeled document corpora.
Pre-training captures foundational knowledge of semantic and syntactic relationships in baseline models useful for fine-tuning on specific tasks downstream with fewer examples needed.

The models used in our work build on Morris et al. \cite{morris2023using} by adopting the Longformer \cite{Beltagy:2020:Longformer} architecture.
Longformers are an embedding transformer model based on RoBERTa \cite{Liu:2019:RoBERTa} that tokenize a sequence of words as input, replace each token with a $768$-dimensional embedding in a semantic vector-space, sum the embeddings with positional embeddings to encode relative position information, and finally feed the embedding matrix into a neural network model.
The model comprises 12 attention layers in which the input embedding matrix is transposed and multiplied by itself to form similarity metrics pairwise between embeddings, or \textbf{attention weights}.
For most BERT-style models, each token at each attention layer is attended to by each other token, resulting in higher demands for computation as the input sequence length increases.
As a result, these models have a limited max sequence length typically around $512$ tokens.
However, providing additional context by including source text as input during LLM training increased performance compared with training on summaries alone \cite{morris2023using}.
To solve input sequence length limitations, Longformer assigns \textbf{global attention} to one or more tokens and a \textbf{sliding attention window} which moves across the text.
Tokens in global attention are attended to by every other token while tokens in the sliding attention window are attended to only by other tokens within the window.
As a result of this design, Longformer can accept much longer texts, up to $4096$ tokens, while remaining computationally efficient.
To increase model interpretability, \textit{iScore} addresses novel challenges around visualizing the sliding window, global attention, and the increased number of tokens and attentions at scale.

\subsection{Interpreting ML Using Visual Analytics}
\label{sec:VA_LLM_interpretability}
Visual analytics is an increasingly popular approach for analyzing and interpreting machine learning (ML) models \cite{Hohman:2019:VAinDeepLearning, Liu:2017:AnalysisofMLwithVASurvey}.
We build on a rich history of visual analytics tools for understanding educational data \cite{Emmons:2017:MOOCVAReview, Vieira:2018:VAEduDataReview, Zhang:2022:VisOnlineLearningReview}, drawing inspiration specifically from systems that seek to visually explain ML-powered educational technologies.
In this space, Mubarak et al. built a system for visualizing patterns of learner interactions with videos in Massive Open Online Courses (MOOCs) to help explain predictions of learners' weekly performance from the interaction data using deep learning models \cite{Mubarak:2021:VAClickstreamDLMOOC}.
Chen et al. engaged a user-centered design process with instructors and ML researchers to design and develop DropoutSeer \cite{Chen:2016:DropoutSeer}, a visualization system for visually explaining ML predictions of learner dropout in MOOCs.
Garcia-Zanabria et al. utilized a similar approach to visually explain predictions of learner dropout using counterfactual explanations in their system, SDA-Vis \cite{Garcia-Zanabria:2022:SDA-Vis}.
Zhang et al. leverage deep learning models (CNN and LSTM) to predict learner dropout and visualize counterfactual explanations using DropoutVis \cite{Zhang:2023:VADropoutCounterfactual}.
A common theme that cuts across these tools is a focus on human-in-the-loop workflows that enable data scientists like learning engineers to inject human intuition and domain expertise into the iterative process of training and validating model performance \cite{Wang:2021:HITLNLP}, which \textit{iScore} makes heavy use of.

To increase transparency for learning engineers, we aim to visualize LLM performance across multiple interpretability methods.
One method is directly visualizing a model's internal architecture as a form of explanation \cite{Strobelt:2019:Seq2seqVis, Wang:2020:CNNExplainer}.
With transformers, several tools focus on visualizing the internal prediction process of transformers as changes in attention weights across each layer and head of a model \cite{Vig:2019:BertViz, Hoover:2020:ExBERT, Derose:2020:AttentionFlows, Wang:2021:Dodrio}.
However, there is debate as to whether attention weights in transformers can be used as a source of interpretation for model performance \cite{Clark:2019:AnalyzingBERTAttention, Jain:2019:AttentionIsNotExplanation, Wiegreffe:2019:AttentionIsNotNotExplanation, Atanasova:2020:ExplainingUsingWeights}.
Alternatively, structuring visual comparison between variations in model inputs and resulting outputs presents a more flexible and model-agnostic analysis approach.
For example, VizSeq is a visual analysis toolkit for interactively evaluating language model task benchmarks \cite{Wang:2019:VizSeq}.
Other tools present a visual analytics workflow to analyze changes in language model weights under various task-specific scenarios \cite{Tenney:2020:LIT, Geva:2022:LMDebugger}.
By combining internal and external interpretability techniques in \textit{iScore}, we seek to give users more control and flexibility over how they interact with the models, increasing understanding of model performance using alternative perspectives \cite{Liao:2023:AITransparencyLLMs}.

\section{Design Process}
\label{sec:design}

\begin{figure*}[!t]
  \centering
  \includegraphics[width=\linewidth]{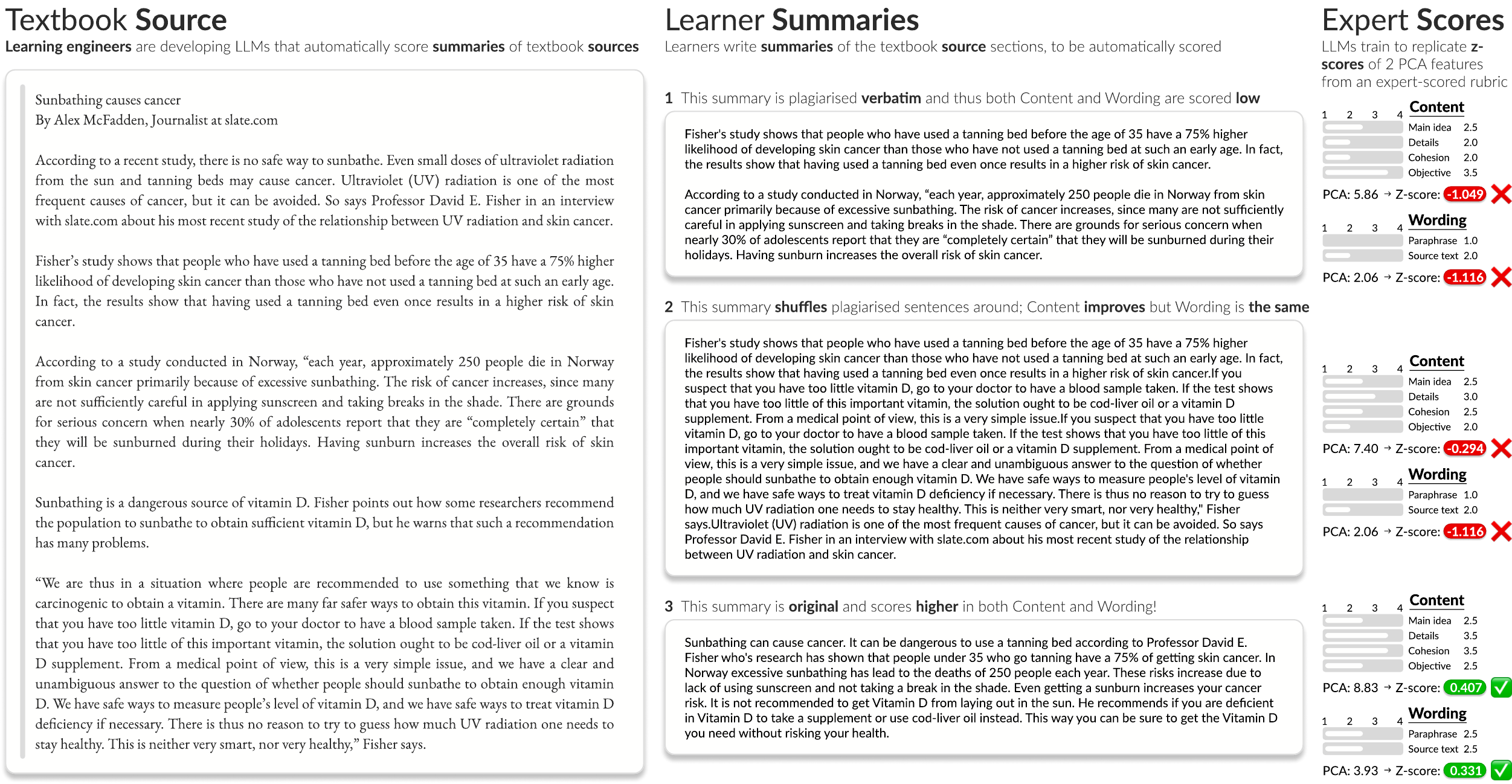}
  \caption[]{%
    An example of a textbook source, learner summaries and expert scores used to train the LLMs visualized in \textit{iScore}.
    Content and Wording LLMs each assign a continuous score that represents components of an analytic rubric.
    Learning engineers seek to characterize how changes in scores relate to differences in summaries via comparison.
    \textit{iScore} provides inputs for multiple summaries per source and visualizes their predicted scores simultaneously in context of the ``ground truth'' training data.
  }%
  \Description{%
    Three columns of text boxes.
    The first column shows an excerpt from a textbook in a text box.
    The second column shows three rows, each with a differently written summary of the textbook excerpt in a text box.
    The third column shows three rows, each with a bar chart of the scores that the summaries received when graded.
  }%
  \label{fig:llm_data}
\end{figure*}

Our goal in this work is to build a visual analytics system that enables data scientists in the learning analytics community, i.e. learning engineers, to interpret how large language models (LLMs) automatically score summary writing, helping them calibrate their trust in these models before deploying them.
To do this, we engaged in user-centered design methodologies including contextual inquiry, rapid prototyping, and design iteration.
Our team comprised all authors and included visualization experts in human centered computing and interaction design as well as learning engineers with expertise in natural language processing (NLP) and developing learning tools.
Over the course of several months, we worked together in multiple virtual formative sessions to develop a shared understanding of the pain points in maintaining a human-in-the-loop workflow for monitoring and improving LLMs used for automatic summary scoring.

We first describe the workflow of our collaborators and their development of summary scoring LLMs (Sect.~\ref{sec:design_background}) presented as examples throughout the paper.
We then summarize the key design challenges and user tasks (Sect.~\ref{sec:design_challenges}) for interpreting how these models are performing that our visual analytics solution addresses.

\subsection{Background: Summary Scoring LLMs}
\label{sec:design_background}
In this paper, we demonstrate \textit{iScore} using summary-scoring LLMs that our collaborators specifically developed for \textit{iTELL}.
In Sect.~\ref{sec:case_study}, we present a case study with a learning engineer from our team who used \textit{iScore} to improve the accuracy of LLMs in \textit{iTELL} by three percentage points.
By scaffolding evaluation around pairs of source and summary text, \textit{iScore} presents a new workflow for learning engineers to interact with summary scoring LLMs.

\subsubsection{iTELL: Textbooks That Score Summaries}
\label{sec:iTELL}
Computational advances at the intersection of artificial intelligence (AI) and natural language processing (NLP) have catalyzed interest in developing intelligent textbooks that present adaptive ``smart'' functionalities to learners such as personalized feedback \cite{brusilovsky2022return}.
To help instructors quickly and simply build intelligent digital textbooks at scale, our collaborators developed \textit{iTELL} (Intelligent Texts for Enhanced Lifelong Learning) as a framework for converting static text into a dynamic interactive web application with minimal labor and technical expertise required from content creators.
A key intelligent feature built into \textit{iTELL} is having learners summarize the section that they have read at the end of each textbook section.
\textit{iTELL} then provides formative feedback on the \textbf{Content} and \textbf{Wording} of these summaries using two LLMs, one for each facet.

\subsubsection{Sources, Summaries and Scores}
\label{sec:llm_data}
Our collaborators trained two BERT-style Longformer \cite{Beltagy:2020:Longformer} models that assign \textbf{Content} and \textbf{Wording} scores to \textbf{summaries} of textbook \textbf{source} content.
After the case study, they trained two more models, \textbf{Content (global)} and \textbf{Wording (global)}.
We evaluate all four models in this paper.

Fig.~\ref{fig:llm_data} shows how scores assigned by the LLMs can vary across revisions of three different summaries of the same textbook source.
The scores are defined during LLM training based on a $1-4$ scaled analytic rubric with six criteria: (1) main idea, or to what extent the summary captured the main idea of the source; (2) details; (3) cohesion, or how well the summary was rationally and logically organized; (4) objective language, or reflecting the point of view of the source; (5) paraphrasing, or avoiding plagiarism by paraphrasing the original material; and (6) language beyond the source, or how well all relevant details were included in the summary.
Training samples were first expert-scored according to the rubric.
Then, to reduce the dimensionality of the LLM output, the rubric criteria for each training sample were distilled using a principal component analysis (PCA) into two components.
Content comprises main idea, details, cohesion and objective language, while Wording comprises paraphrasing and language beyond the source.
The two PCA scores for Content and Wording were then z-score normalized.
In this way, the scores which the LLMs are trained to assign now represent the standard deviation of the scored summary from the mean score along a single dimension, transforming the rubric into a more interpretable and useful result.
These scores are then used as a target label in the training data for each respective model.

The models were fine-tuned on $4690$ different training summaries written on $101$ different source texts, enabling them to adapt to source/summary combinations on different topics with minimal to no adjustments needed.
The training task was modeled as a sentence classification task \cite{Wang:2018:GLUE}, where each model tries to predict a continuous score (i.e. the Content or Wording score) as a label.
In a typical multi-class classification task, the output is composed of multiple labels, each of which receive a logit score.
During training, the model outputs are compared against one-hot encoded target outputs and loss is computed using cross-entropy.
During inference, the max logit score is chosen as the predicted label. 
When predicting a continuous score, however, only a single label is used and the logit score of that label is interpreted as the predicted score. 
Instead of cross-entropy, mean squared error (MSE) was chosen as the loss function because MSE penalizes larger errors more than smaller errors, thus encouraging closer alignment with the predicted labels. 
This method is commonly used in tasks that require the prediction of a continuous label \cite{leonardi2020multilingual, ormerod2022mapping, steimel2020towards}.
The models were trained for six epochs with a batch size of eight and a learning rate of $3e$-$5$.

\begin{table}
  \caption[]{%
    Design Challenges (C) and User Tasks (T)
  }%
  \label{tab:challenges_tasks}
  \begin{tabular}{rl}
    \toprule
      \textbf{C1} & Characterizing how summaries and scores are related \\
    \midrule
      \textbf{T1} & Stratify scoring across several facets of writing \\
      \textbf{T2} & Compare multiple summaries and scores at once \\
      \textbf{T3} & Use expert-scored writing samples as a reference \\
    \midrule
      \textbf{C2} & Comparing differences between versions of summaries \\
    \midrule
      \textbf{T4} & Highlight how revisions affect assignment scores \\
      \textbf{T5} & Investigate multiple different types of revisions \\
      \textbf{T6} & Track the provenance of both the writing and scores \\
      \textbf{T7} & Select and view previous versions of summaries \\
    \midrule
      \textbf{C3} & Understanding model parameters and behaviors \\
    \midrule
      \textbf{T8} & Compare variations in model inputs to model outputs \\
      \textbf{T9} & Examine the internal model weights (i.e. attention) \\
    \midrule
      \textbf{C4} & Bridging global and local model behavior \\
    \midrule
      \textbf{T10} & Summarize interactions between all tokens at once \\
      \textbf{T11} & Compare token, layer and head interactions \\
      \textbf{T12} & Drill down to interactions at specific layers and heads \\
    \bottomrule
  \end{tabular}
\end{table}

\subsection{Design Challenges and User Tasks}
\label{sec:design_challenges}
From our formative sessions, we synthesized several key design challenges \textbf{(C)} and user tasks \textbf{(T)} summarized in Table~\ref{tab:challenges_tasks} and integrated into our descriptions of the system (Sect.~\ref{sec:system}), usage scenarios (Sect.~\ref{sec:usage}) and evaluation (Sect.~\ref{sec:evaluation}) throughout the rest of the paper.

\medskip
\noindent\textbf{C1  }
\textbf{Characterizing the relationship between summaries and scores.}
Summaries balance multiple facets of writing quality (text length, cohesion, paraphrasing, details, etc.), making it difficult to automatically assign a single graded score encapsulating all aspects of importance.
To address this, learning engineers first seek to model each of these facets individually by \textit{stratifying scoring along several dimensions, using multiple scoring models} \textbf{(T1)}.
This can help language models automatically assign more objective scores to specific facets such as plagiarism or grammar.
With multiple scores, engineers can then begin to interpret and improve model performance by \textit{comparing multiple summaries and scores simultaneously} \textbf{(T2)}.
Fig.~\ref{fig:llm_data} provides an example of how differences in summaries (e.g., syntactic and semantic) can lead to differently assigned scores.
To further enrich comparison, engineers also \textit{use expert-scored writing samples as a reference} \textbf{(T3)}.
In the context of our Content and Wording summary scoring models, these expert-scored samples are used as training data.
Our interface should enable users to compare multiple summaries and scores at once, as well as in reference to the "ground truth" expert-scored training data.

\medskip
\noindent\textbf{C2  }
\textbf{Comparing differences between versions of summaries.}
If a summary's scores are too low, revisions need to be made that address the factors that the models base their assessment on.
It is critical to give learning engineers methods that \textit{highlight how revisions affect automatic score assignments} \textbf{(T4)}.
Engineers are also interested in \textit{investigating multiple types of revisions that can include spelling and grammar, paraphrasing, and keywords} \textbf{(T5)}.
Together, these methods can reveal insights into how the language models are working and provide engineers with useful criteria for feedback to learners as they write their summaries.
As revisions accumulate, our system should \textit{track the provenance of both the writing and scores} \textbf{(T6)}, and allow users to \textit{select and view previous versions of summaries} \textbf{(T7)}.
This compliments the need for comparison of multiple summaries and scores simultaneously in Fig.~\ref{fig:llm_data} by adapting the comparison to multiple versions of the same summary.

\medskip
\noindent\textbf{C3  }
\textbf{Understanding model parameters and behaviors.}
Using a deep learning approach for automatically assigning scores introduces opaqueness into understanding model behavior \cite{Liao:2023:AITransparencyLLMs}, making it difficult to support learning engineers in closing the NLP loop of model development \cite{Wang:2021:HITLNLP}.
In the context of scoring written summaries, engineers grapple with two difficult tasks when interpreting transformer-based language models specifically.
The first is \textit{understanding how variations in model inputs can lead to changes in model outputs} \textbf{(T8)}.
For example, counterfactual explanations \cite{Wallace:2019:TrickMeIfYouCan, Garcia-Zanabria:2022:SDA-Vis, Zhang:2023:VADropoutCounterfactual} are often used to explore model behavior and robustness by systematically varying and comparing different types of revisions.
The second is \textit{interpreting the internal model weights (i.e. attention in transformers)} \textbf{(T9)} to understand what the model is capturing about summaries.
This could enable engineers to identify emergent representations of semantic knowledge and lexical structure that contribute to certain model scores \cite{Rogers:2020:Bertology}; e.g., identifying salient interactions between specific character spans across model layers and heads.
Providing both external and internal model probing methods can guide users in discovering useful insights to help them close the NLP loop of model development.

\begin{figure*}[!t]
  \centering
  \includegraphics[width=\linewidth]{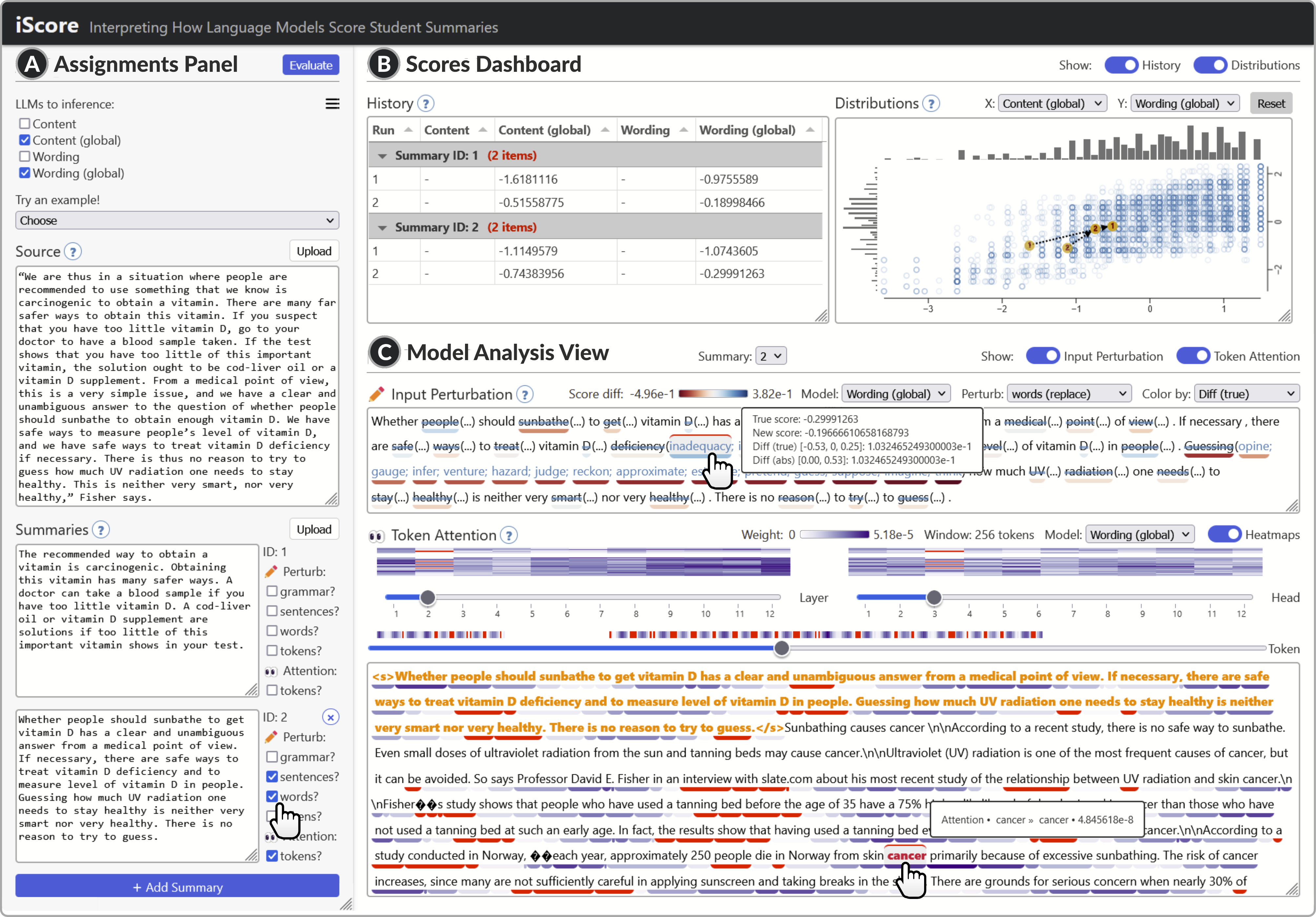}
  \caption[]{%
    \textit{iScore} visualizes multiple LLM-scored writing samples to help learning engineers interpret model performance.
    Above, a learning engineer interprets how two plagiarized summaries are scored across two runs (Sect.~\ref{sec:usage}).
    Users can upload, score and manually revise and re-score multiple source/summary pairs simultaneously in the \textit{Assignments Panel} \textbf{(A)}, visually track how scores change across revisions in the context of expert-scored LLM training data in the \textit{Scores Dashboard} \textbf{(B)}, and compare model weights between words across model layers/heads, as well as differences in scores between automatically revised summary perturbations, using two model interpretability methods in the \textit{Model Analysis View} \textbf{(C)}.
  }%
  \Description{%
    The iScore web interface.
    From top to bottom, the page consists of a small full-width header and a main content area with a smaller column on the left and larger column on the right.
    The left column labeled (A) shows the Assignments Panel with a source and two summaries loaded into text boxes.
    The right column has two rows.
    The top row labeled (B) shows the Scores Dashboard with a table of scores and a scatter plot of training data scores and scores from the table.
    The bottom row labeled (C) shows two model interpretability visualizations; input perturbation and token attention.
    Input perturbation shows summary text in a text box with some individual words underlined.
    Token attention shows two heat maps side-by-side labeled layer and head with sliders below them.
    Below that, a horizontal rug plot of one of the heat map columns is shown with a slider below it.
    Finally, at the bottom, summary text is plotted in a text box with some individual words underlined.
  }%
  \label{fig:system}
\end{figure*}

\medskip
\noindent\textbf{C4  }
\textbf{Bridging global and local model behavior.}
In addition to existing challenges interpreting deep learning model behavior, learning engineers also face a scale issue, as the models used in \textit{iTELL} can score textbook sources and summaries thousands of words (tokens) in length.
This makes summarizing the sheer scale of tokens and attentions while helping users make sense of the data difficult.
To address this, engineers seek methods to \textit{summarize the interactions between all tokens in the same interface} \textbf{(T10)}.
Bridging the gap between global and local model behavior can be supported by \textit{comparing interactions between tokens, layers and heads} \textbf{(T11)} while enabling users to \textit{drill down to specific token interactions at specific layers and heads} \textbf{(T12)}.
By aggregating data at multiple levels, our system should reveal subsets of interesting interactions that help engineers make sense of complex model behaviors.

\section{The \textit{\MakeLowercase{i}S\MakeLowercase{core}} System}
\label{sec:system}

Based on our design challenges and user tasks, we developed \textit{iScore}, an interactive visual analytics tool for learning engineers to upload, evaluate, and visualize the results of automatically scoring summary writing using LLMs.
Our system helps users characterize quality in writing samples and enable transparency in LLM performance by tightly integrating three coordinated views.
In the \textit{Assignments Panel} (Sect.~\ref{sec:assignments_panel}), users scaffold the evaluation process around comparing multiple source/summary pairs simultaneously.
After evaluating the pairs, the \textit{Scores Dashboard} (Sect.~\ref{sec:scores_dashboard}) visualizes the provenance of scores in context of LLM training data, while the \textit{Model Analysis View} (Sect.~\ref{sec:model_analysis}) presents two different LLM interpretability methods at multiple levels of abstraction.

\subsection{Assignments Panel}
\label{sec:assignments_panel}
The \textit{Assignments Panel} (Fig.~\ref{fig:system}A) provides multiple source/summary text inputs and model analysis options.
This structure guides users to compare different facets of summary writing in the \textit{Scores Dashboard} and \textit{Model Analysis View} by choosing between multiple model analysis methods.

We scaffold the evaluation of source/summary pairs in several ways.
Users can choose any number of language models they have trained from a list to customize how they evaluate assignments and which models are being compared \textbf{(T1)}.
We also provide a list of example source/summary pairs to help users understand how to start inputting sources and summaries.
Users can then provide a single source, either through copy/paste, upload, or by directly typing and editing in a free text box.
For the given source, users can add any number of summary text boxes with the same input capabilities \textbf{(T2)}, with a list of model analysis option checkboxes for each that will populate the visualizations in the \textit{Model Analysis View} for each summary with a checkbox selected \textbf{(T8, T9)}.
Because each model analysis option can be computationally expensive, users can toggle which options they would like to run independently for each summary. 
We explore these options in detail in Sect.~\ref{sec:model_analysis}.

Source/summary pairs are scored in real time using an API that interfaces with a Python Flask server running PyTorch implementations of the language models.
We use the HuggingFace Transformers \cite{Wolf:2020:HuggingfaceTransformers} API to load models and perform sequence classification by combining each summary with the source using a separator token.
Users can rapidly generate multiple test cases across several summary variations for a given source and evaluate them simultaneously, saving time and improving efficiency.

\begin{figure}[!t]
  \centering
  \includegraphics[width=\linewidth]{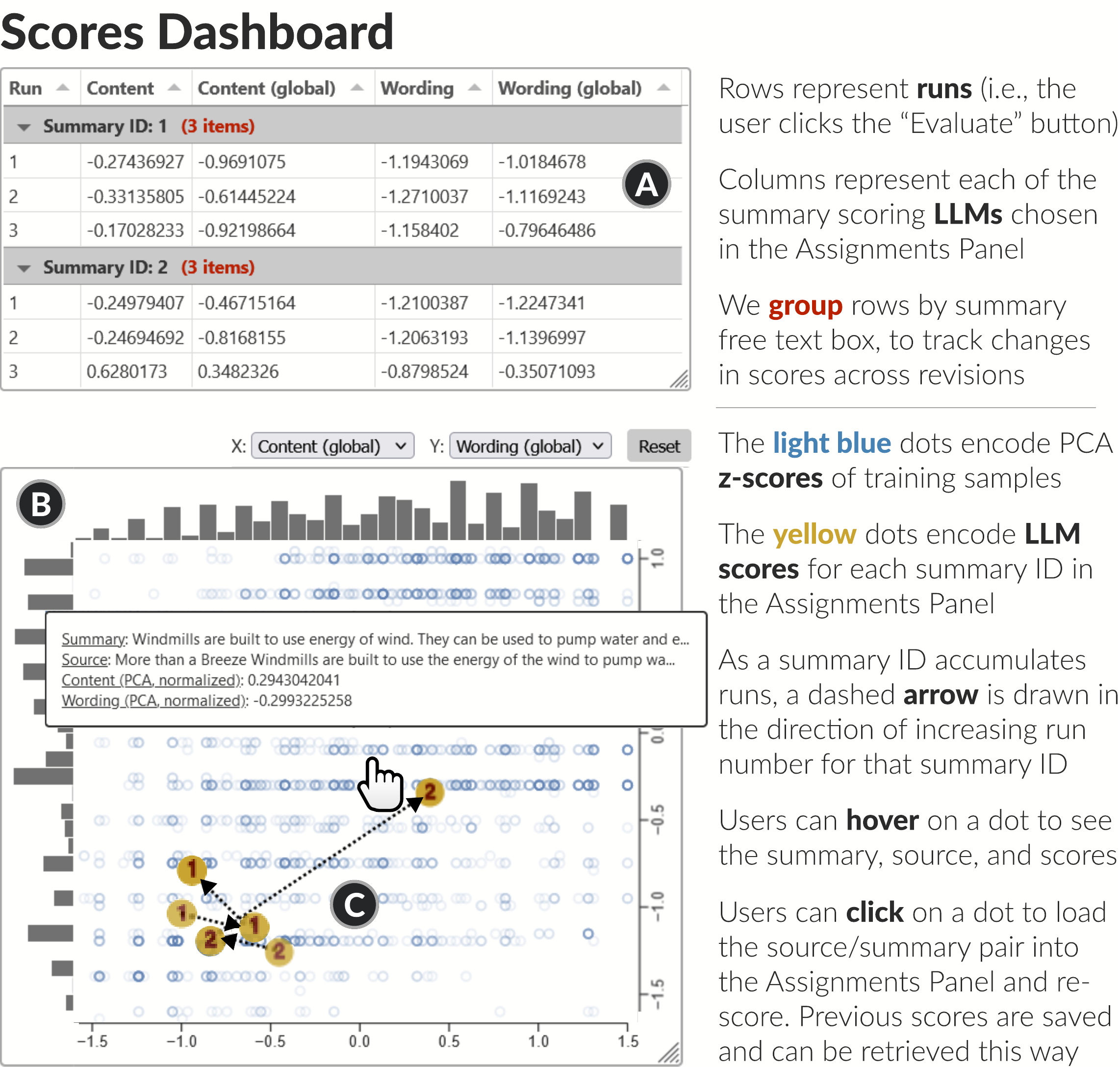}
  \caption[]{%
    Breakdown of the \textit{Scores Dashboard}.
    The table and scatter plot help users compare variations on summaries by tracking how scores change across manual revisions.
  }%
  \Description{%
    A diagram of the Scores Dashboard with text box call-outs.
    On the left, the history table is shown above the distributions scatter plot.
    On the right, text box call-outs describe how to interpret and interact with the visualizations.
  }%
  \label{fig:scores_dashboard}
\end{figure}

\subsection{Scores Dashboard}
\label{sec:scores_dashboard}
After scoring the source/summary pairs, we provide an overview of model scores for each summary in the \textit{Scores Dashboard} (Fig.~\ref{fig:system}B).
We visualize the provenance of score data to help users gain insight into how changes in summaries affect model scores across runs of the model at different levels of aggregation.

Each time the user scores a set of source/summary pairs, e.g., when revising a summary to see how scores change, we visualize the provenance of raw summary scores as a table (Fig.~\ref{fig:scores_dashboard}A) \textbf{(T6)}.
Runs of the models are numbered.
Rows represent a single source/summary pair, labeled by run and grouped by summary free text box.
Columns display the scores for each LLM chosen in the \textit{Assignments Panel}.
In this way, we facilitate rapid comparison across both runs for a single summary, as well as across summaries for the same run.
For example, users can easily determine if scores are increasing/decreasing when a single summary is revised or discover which summaries are performing better compared with others across runs.
Users may also want to understand how summary scores from the \textit{Assignments Panel} compare with the ground truth model training examples (Sect.~\ref{sec:llm_data}), helping them gain additional context into why a particular summary received a certain score.
To facilitate this, we visualize training example scores directly as a scatter plot with overlapping blue dots (Fig.~\ref{fig:scores_dashboard}B) \textbf{(T3)}.
We then plot the scores of the current summaries in the \textit{Assignments Panel} as yellow dots overlaid on top of the training example score distributions (Fig.~\ref{fig:scores_dashboard}C).
To dig even deeper, our collaborators requested the ability to re-score the training examples, helping them better understand the differences between training and testing on the ground truth data set.
In response, we allow users to click on any dot in the scatter plot to load that source/summary example into the \textit{Assignments Panel} and score it.
Scores are saved and can be loaded on the fly by clicking on the corresponding dot \textbf{(T7)}.

As users explore the history and distributions of scores, we provide several interactions and useful overlays.
To help users gain additional context and insight, we visualize the provenance of summary scores in the \textit{Assignments Panel} by plotting each run of a summary as a dot numbered by the run \textbf{(T6)}.
A dashed arrow line indicates the direction that the score is moving for that summary between runs.
To solve issues with occlusion, we made each training example dot mostly transparent, revealing patterns in training example score distribution density where dots overlap.
We also explicitly plot the distributions of training example scores as bar chart histograms along the scatter plot axes \textbf{(T3)}.
To see details on demand, users can hover on dots to reveal a snippet of the source/summary pair text for that example as well as the numeric scores.
They can also hover on bars to view the bin size and range.

\subsection{Model Analysis View}
\label{sec:model_analysis}
The \textit{Model Analysis View} (Fig.~\ref{fig:system}C) allows users to interpret and gain deeper insights into how the LLMs are scoring each source/summary pair in the \textit{Assignments Panel}.
We do this by leveraging two broadly effective techniques for interpreting deep learning models at multiple levels of text aggregation: (1) visualizing the results of input perturbation on model outputs in the \textit{Input Perturbation} plot (Sect.~\ref{sec:input_perturbation}); and (2) visualizing attentions of transformer-based language models between tokens across every model layer and head in the \textit{Token Attention} plot (Sect.~\ref{sec:token_attention}).
We visualize these methods applied to a single source/summary pair at a time; users can switch between pairs using a drop-down menu.

\begin{figure}[!t]
  \centering
  \includegraphics[width=\linewidth]{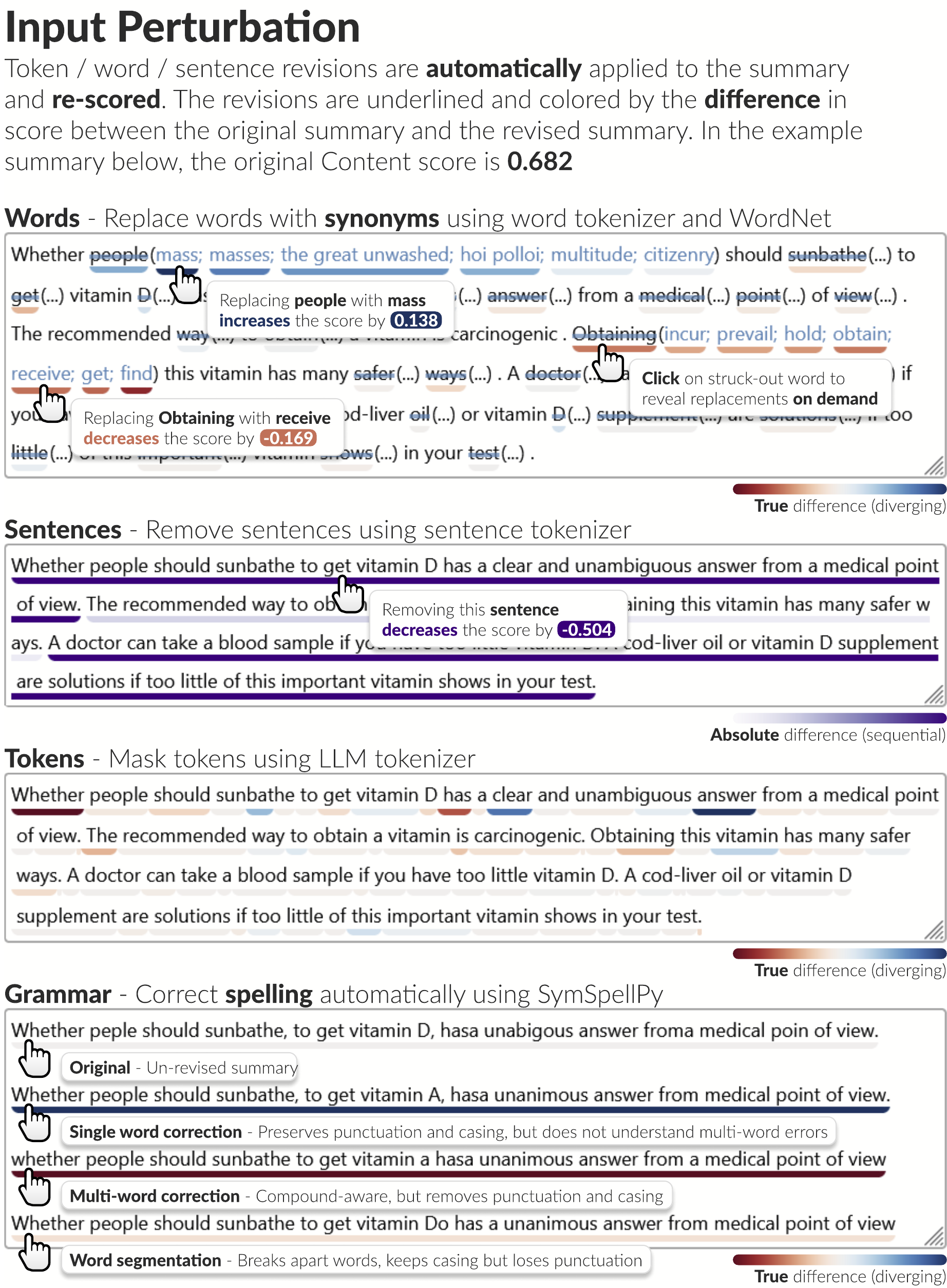}
  \caption[]{%
    Breakdown of the \textit{Input Perturbation} visualization.
    Multiple perturbation methods help users test hundreds of different kinds of revisions at scale by automatically applying and re-scoring summaries for them.
  }%
  \Description{%
    A diagram of the Input Perturbation visualization with text box call-outs.
    Four full-width rows of the visualization are shown, each with a different perturbation method applied to a summary.
    Text box call-outs explaining the visual encodings are overlaid on the visualizations with mouse cursors pointing to where users would hover and click.
  }%
  \label{fig:input_perturbation}
\end{figure}

\subsubsection{Input Perturbation}
\label{sec:input_perturbation}
Perturbing input parameters during model inference has been widely used to explore the inner workings of black-box AI models, particularly in scenarios where complex model architectures make it difficult to trace explainability through the entire model from input to output \cite{Ribeiro:2016:LIME, Lundberg:2017:SHAP}.
In the case of using transformer-based language models, we created four summary text perturbation options in the \textit{Assignments Panel} \textbf{(T5)}.
The \textbf{grammar} and \textbf{words} perturbation options both replace word-level spans by fixing the spelling and using synonyms respectively, while the \textbf{sentences} and \textbf{tokens} perturbation options mask sentence-level and token-level spans from the summary respectively.
After automatically perturbing each summary, we inference each LLM requested in the \textit{Assignments Panel} and visually compare the new summaries' scores with the original, unperturbed summary score \textbf{(T4)}.
These perturbation options mirror typical methods of evaluating summary writing suggested by our domain expert collaborators, and were developed and tested as a proof of concept.

\medskip
\noindent\textbf{Data.  }
The \textbf{words} option replaces all word-level spans that are not English stop words or punctuation with synonyms in each summary, giving users an overview of which words have a strong effect on each summary's score when replaced and if semantically similar word replacements could improve the score.
We perform replacement by first identifying all word-level spans using a treebank tokenizer, then looking up the synonyms of each word and returning each synonym's lemma.
Words are replaced by their synonyms' lemmas and a new summary score is computed for each lemma, resulting in a list of scores for each synonym replacement.

Both the \textbf{sentences} and \textbf{tokens} options apply masking to the summary at the sentence-level and token-level, respectively.
Masking spans attempts to replicate saliency in model interpretability, which assigns a score to each input based on the strength of its contribution to the final output.
In \textit{iScore}, we use the difference in score when masking spans as a proxy for the importance of the span to the unperturbed summary score.
We use a sentence boundary detection tokenizer to locate each sentence-level span in the summary, while we use each LLM's tokenizer used for inference to extract token-level spans that correspond with the tokens in the \textit{Token Attention} plot (Sect.~\ref{sec:token_attention}).
As with the \textbf{words} option, we compute a new summary score for each masked span.

Finally, the \textbf{grammar} option attempts to automatically correct the spelling of each summary.
This allows users to quickly explore how the quality of the grammar affects each summary score.
We first perform a correction on the entire summary then query each model with the corrected summary, to get a single score.
We implemented three correction variations for comparison.
The first performs single word spelling correction using lookups on individual word spans generated from a treebank tokenizer.
This allows us to accurately preserve the structure of the sentence (i.e. the casing and punctuation) but may miss context-dependent spelling suggestions that are likely to occur.
The second performs compound-aware multi-word spelling correction on the entire summary.
In contrast with single word spelling correction, the accuracy of recognizing and fixing multi-word errors is likely to be higher yet the structure of the sentence is lost in the process.
The third attempts word segmentation to divide multi-character spans into their constituent words, preserving casing but ignoring punctuation.

\medskip
\noindent\textbf{Visualizations.  }
To visualize the results of perturbation, we plot the entire summary text in a text box and underline text spans (token-, word-, or sentence-level) that correspond with the perturbation options chosen in the \textit{Assignments Panel} (Fig.~\ref{fig:input_perturbation}).
We then encode the difference between each perturbed summary's score and the unperturbed summary's score on a color ramp and apply the color to the text underlines \textbf{(T4)}.
This encoding scheme allows users to quickly identify and compare the length and effect size of each span in the perturbed summary with the unperturbed summary score.
Users can choose between a red-to-blue diverging color ramp representing the true negative-to-positive difference in scores, or a transparent-to-purple linear color ramp representing the absolute value of the difference in scores.
We display a legend for the underline colors above the plot.
To get details on demand, users can hover on an underlined span at any level to get the exact difference values.
Finally, we provide drop-down menus to let users switch between models, perturbation options, and color ramps.

The summary text and underlines are plotted differently based on the perturbation option (Fig.~\ref{fig:input_perturbation}) \textbf{(T5)}.
For the \textbf{words} option, we implemented a word replacement view common in most text editors with markup capabilities.
Replaced words have a strike-through decoration and are followed by ellipses, denoting a replacement was made for that word.
This gives users an overview at a glance of which words have hidden synonym replacements that affect the unperturbed summary score the most.
Users can click on any word with replacements to show/hide that word's synonyms on demand.
Each synonym is shown in a list with a text underline colored by the difference in summary score when using that synonym in place of the original word.
The replaced word is also underlined; the underline is colored by the maximum signed magnitude of score differences for all synonym replacements of that word.
For the \textbf{sentences} and \textbf{tokens} options, we underline each sentence-level and token-level span respectively with the color representing the difference in summary score when that span was masked during inference.
This allows users to quickly identify spans with strong effects on the final model output.
For the \textbf{grammar} option, we plot and underline the entirety of each correction variation of the unperturbed summary in a single color representing the difference in the corrected summary's score.
Users can then compare changes in grammar and scores side-by-side between corrections.

\begin{figure}[!t]
  \centering
  \includegraphics[width=\linewidth]{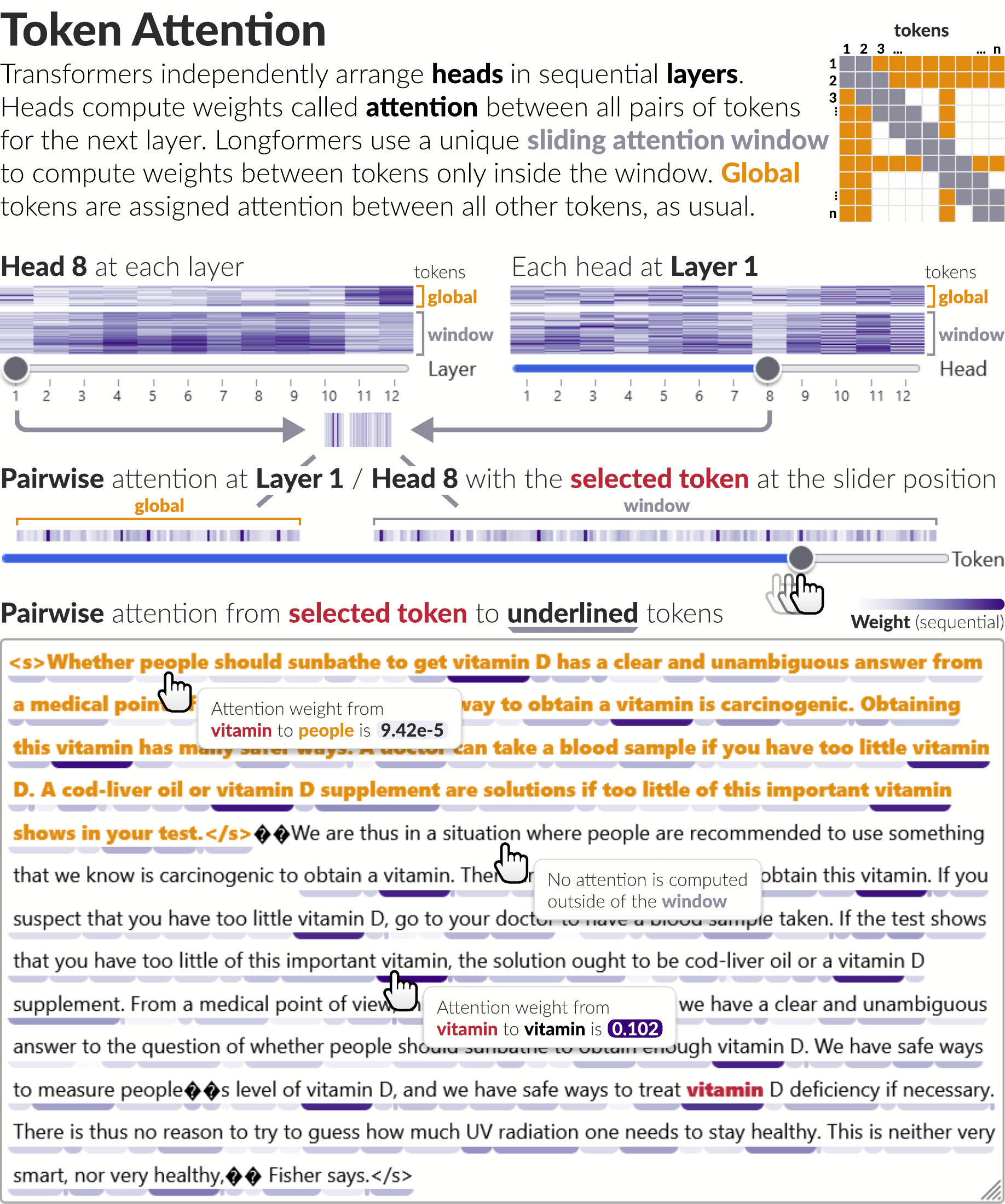}
  \caption[]{%
    Breakdown of the \textit{Token Attention} visualization. 
    The combination of heat maps, rug plot and text underlining helps users make sense of complex model behaviors by keeping them in the loop at multiple levels of abstraction.
  }%
  \Description{%
    A diagram of the Token Attention visualization with text box call-outs.
    Three full-width rows of the visualization are shown.
    The first row shows heat maps with sliders below them and annotations pointing to tokens with global versus window attention.
    Below that, a column of the heat map is rotated 90 degrees counterclockwise and drawn as a rug plot on top of a slider with a mouse cursor shown dragging the slider.
    Finally, below that, summary text is plotted in a text box and underlined.
    Text box call-outs explaining the visual encodings are overlaid on the visualizations with mouse cursors pointing to where users would hover and click.
  }%
  \label{fig:token_attention}
\end{figure}

\subsubsection{Token Attention}
\label{sec:token_attention}
Complementing our approach of evaluating models externally, we also visualize the internal multi-headed attention mechanisms underlying our transformer-based language models.
By visually connecting attention to the semantic and syntactic structure of our source/summary pairs, we aim to reveal patterns and associations in the text that are strongly contributing to the final model output.
This can help users gain deeper insights into how different training procedures and training data lead to prioritizing different spans between layers/heads, and compare how attentions and these priorities change between models.

\medskip
\noindent\textbf{Data.  }
Prior work \cite{Wang:2021:Dodrio, Vig:2019:BertViz} has successfully explored using graph and matrix visualization techniques for visualizing pairwise attention between tokens for models with smaller sequence lengths ($<256$ tokens).
However, a major challenge in scaling attention visualization for our models is the size of the sliding context window ($256$ or $512$ tokens) compared to the input sequence length (up to $4096$ tokens) for Longformers \cite{Beltagy:2020:Longformer}.
Longformer assigns \textbf{global attention} to one or more tokens and a \textbf{sliding attention window} which moves across the text.
Tokens in global attention generate attention to every other token, while tokens in the sliding attention window only generate attention to other tokens within the window.
This creates issues in visualizing the sheer number of tokens and attentions both locally (within the context window) and globally (for all tokens), while helping users make sense of the data.
For example, using a $256$-token sliding context window around each token, a source/summary pair with $700$ tokens ($\approx500$ words, such as a 5-paragraph source and 1-paragraph summary) results in a 4-dimensional matrix of $700\times256\times12\times12\approx26$ million attentions between each token for a model with $12$ layers and $12$ heads in each layer.
If we consider global attention between every token, this number increases to $700\times700\times12\times12\approx70$ million attentions.
Textbook sources and summaries in \textit{iTELL} regularly exceed $4000$ tokens ($>150$ million attentions).
Because our models inference using both the source and summary, we considered designs that allowed us to scale up to $4096$ token attentions visualized at once.

\medskip
\noindent\textbf{Visualizations.  }
To help users discover higher-level patterns and drill down into specific layers and heads, we provide an overview of attention from the selected token to all other tokens using heat maps (Fig.~\ref{fig:token_attention}) \textbf{(T10)}.
We plot two different heat maps as uniform grids of all other tokens around the selected token arranged vertically, and each layer (i.e. the Layer heat map) or head (i.e. the Head heat map) arranged horizontally.
Cells are colored by the attention weight from the selected token to the other token (row) at that layer or head (column).
A horizontal single-ended range slider is aligned below each heat map and allows users to select a particular layer/head combination for viewing in the text box described below.
As the range slider for the Layer or Head heat map is dragged, the other heat map not being dragged will update.
For example, if Layer 2 is selected in the Layer heat map, then the Head heat map will show an overview of attentions across all the heads of Layer 2.
This allows users to quickly pan through layers and heads to gain insights into how attention changes in other heads and layers, respectively.

Below the heat maps, we allow the user to drill down and see all pairwise attentions between a selected token and all other tokens at the current layer/head selected in the heat maps (Fig.~\ref{fig:token_attention}) \textbf{(T11)}.
To do this, we position a rug plot above a horizontal single-ended range slider that represents the index of the selected token, and visualize the attention weight from the selected token to all other tokens at each index.
Similar to the \textit{Input Perturbation} plot, we then plot both the source and summary text in a text box and underline each token \textbf{(T12)}.
Each token underline is colored by the attention weight from the selected token to that token.
In this way, we address issues of scale as the number of tokens increases by leveraging basic text formatting to help users read the text naturally while comparing attention weights at the same time.
Users can manually click on any token in the text box to update the heat maps, rug plot, and text underlines with updated attention weights from the newly selected token.
They can also use the range slider to quickly pan through tokens and see patterns at the dataset-level.

We provide several interactions and visual embellishments.
A single linear transparent-to-purple color ramp is used to represent attention weight for the heat maps, rug plot, and text underlines.
Where attention is zero (not missing) between tokens, we color cells, strips, and underlines in bright red to make these outliers visually distinct.
We display a legend for the heat maps above the plot.
Because our models use a sliding context window, in all plots the attention weights for tokens outside of the context window are not plotted, and attentions from the selected token to all global tokens are always shown.
In the text box, global tokens have a golden stroke applied and the selected token has a red stroke applied, to visually distinguish them from other tokens.
To get details on demand, users can hover over any cell, strip, or token and get the token text and exact attention weight value.
Finally, we provide drop-down menus to let users switch between models.

\subsection{Implementation}
\label{sec:implementation}
\textit{iScore} is an open-source\footnote{\textit{iScore} models and code: \url{https://github.com/AdamCoscia/iScore}} web app built using D3.js \cite{Bostock:2011:D3}.
We implemented our input perturbation methods in Python using symspellpy\footnote{symspellpy code: \url{https://github.com/mammothb/symspellpy}} for revising grammar and NLTK's \cite{Bird:2009:NLTK} WordNet interface for word synonym replacement and sentence masking.
\textit{iScore} uses HuggingFace Transformers API \cite{Wolf:2020:HuggingfaceTransformers} to load Longformer pre-trained models and process all source/summary pairs in real time.

\section{Usage Scenario}
\label{sec:usage}

A learning engineer is manually revising two plagiarized summaries of a textbook source in \textit{iTELL} and comparing their LLM-assigned scores using \textit{iScore}.
They seek to: (1) see how they can trick the models into giving the summaries higher scores by reordering and rephrasing sentences; and (2) determine if modifying or removing certain phrases such as key words and main ideas will affect model scores.
They hope to use their findings to develop feedback for learners writing summaries of textbook sections on how to improve scores, as well as generate additional training examples that ensure the models are resilient to learners trying to ``game'' the system.

They start by using the \textit{Assignments Panel} to upload a source text and two summaries \textbf{(T2)}.
While they expect that the models will score plagiarized summaries low on both Content and Wording, they are unsure how resilient the models are to reordering and rephrasing, and whether changes in Content and Wording scores will differ \textbf{(T1)}.
In the first run, they score the plagiarized text verbatim, resulting in negative scores that indicate the summaries are below average.
In the second run, they reorder the sentences without modifying words.
The history table in the \textit{Scores Dashboard} shows a slight increase in both scores between runs, yet the scores are still negative, a good sign \textbf{(T6)}.
They try a third revision, by paraphrasing each sentence without changing the order of sentences.
Interestingly, the dashed arrow lines in the distributions scatter plot show a stronger increase in Content over Wording.
Finally, they reorder and paraphrase the summaries to test the combined effects.
This time, Content increases further but Wording actually decreases.
If paraphrasing alone improves both Content and Wording while reordering and paraphrasing only improves Content, the learning engineer worries that learners may be able to paraphrase summaries using paraphrasing tools like ChatGPT that could pass.
To verify this pattern, the engineer clicks on a training source/summary pair with similar scores in the distributions scatter plot to load it in the \textit{Assignments Panel} \textbf{(T3, T7)}.
They apply the same reordering and rephrasing, and find it produces a similar score pattern.
By understanding the effects of reordering versus rephrasing on model performance using \textit{iScore}, the learning engineer can augment training samples and re-train new model versions.
Once they trust the re-trained models are robust, they can confidently warn learners that the models are aware of these methods and that summaries should be written from scratch else they will fail.

The learning engineer wants to follow up on their intuition that certain words being rephrased or sentences being removed leads to changes in score.
However, it would be time-consuming and inefficient to manually replace every word and remove every sentence, and certain replacements could be missed.
Using the latest revised versions of the summaries, they choose the sentences and words perturbation model analysis options in the \textit{Assignments Panel} for each summary \textbf{(T5, T8)}.
Viewing the sentences option in the \textit{Input Perturbation}, they are surprised to see that the first sentence tends to drop the score dramatically when removed \textbf{(T4)}.
Our collaborators expand on the potential reason for this result in the expert feedback in Sect.~\ref{sec:feedback}.
Switching to the words option, the engineer finds that more complex synonyms generally increase the Content score, indicating a potential learned association between word complexity and score.
For the purposes of the summary assignment, this is considered a positive learned association.
However, they then check the Wording model and find the same pattern does not hold; synonyms not found in the source tend to lower the Wording score.
This is surprising, since Wording consists of rewarding summaries that go beyond the source text.
The engineer makes two decisions: (1) re-train the models by exposing them to more complex summaries; and (2) give learners feedback that they should generally use a variety of vocabulary that goes beyond the source text, but that does not change the meaning of their sentences.

\section{Evaluation}
\label{sec:evaluation}

Our goal for evaluation was to validate the effectiveness of \textit{iScore} for helping learning engineers understand, evaluate, and build trust in how their large language models (LLMs) automatically score summary writing.
To do this, we first deployed \textit{iScore} with three of the learning engineers (domain experts) with whom we collaboratively designed \textit{iScore} over the course of a month.
We first describe a case study detailing how \textit{iScore} enabled one of the engineers on our team to improve the accuracy of LLMs used in \textit{iTELL} by three percentage points (Sect.~\ref{sec:case_study}).
We highlight which challenges and tasks raised in Sect.~\ref{sec:design_challenges} and Table~\ref{tab:challenges_tasks} they achieved.
We then conducted individual semi-structured interviews ($1-2$ hours long each) with all three engineers on our team and collected feedback \cite{Sedlmair:2012:DesignStudyMethods} on the ways \textit{iScore} generally impacted how they understood, evaluated, and trusted the LLMs used in \textit{iTELL} (Sect.~\ref{sec:feedback}).

\subsection{Case Study}
\label{sec:case_study}
One of the domain experts on our team that we deployed \textit{iScore} with and interviewed is a learning engineer who has expertise in natural language processing (NLP).
He is fairly new to deep learning and wants to improve the LLMs deployed in \textit{iTELL} for automatically scoring summary writing.
To do this, he trains several LLMs using various combinations of parameters and evaluates their performance on sample writing by comparing scores one at a time.
He is interested in ``opening the black box'' of the LLMs and feels the most important use of \textit{iScore} is helping engineers confirm that their model is doing what they want it to do: \textit{``You don't see the model working. You set everything up as carefully as you can but it is all done in the GPU. It's easy to make a mistake, allow information leaks, and there is nothing to `see'. You don't know where you might have messed up.''}

During deployment, he used the \textit{Token Attention} visualization to investigate the relationship between the summary and source text.
Looking at the Content model, he noticed that the attention weights tended to converge in the final model layers to high values only in a small radius around a selected token (i.e., darker purple underlines fading linearly to light purple for tokens farther away) \textbf{(T9, T10)}.
This did not match his mental model of what the LLM should be doing.
Tokens in the source should pay attention to the entire summary in each layer, even when they are far away from the summary.
He wondered whether the LLM was using all available information to determine the final score.
He quickly compared this qualitative pattern with other source/summary pairs as well as with the Wording model using the \textit{Assignments Panel} and confirmed the same insight.
He realized that, \textit{``by default, only the classification token receives attention. By using the [token slider in the] attention visualization, I watched the attention window slide across the text. It was obvious that only one token was getting global attention.''} \textbf{(T11, T12)}
The insight gave him an idea to improve the model, by assigning global attention to the entire summary during training and inference.
After retraining, he found the Content score improved significantly, from $r^2=0.79$ to $r^2=0.82$.
He reflected on this improvement, saying \textit{``It's a simple thing, but the attention visualization confirms that the model now does what the model was supposed to do! I gained an important insight using iScore that changed the model. It was actionable information that led to a better model.''}

Overall, the learning engineer overcame the challenges outlined in Sect.~\ref{sec:design_challenges} by using \textit{iScore} to draw a relationship between summaries and scores \textbf{(C1)}, uncover discrepancies between models and differently scored summaries \textbf{(C2)}, deeply investigate errant model parameters \textbf{(C3)}, and fluidly interact with the visualizations to transition between global and local model behavior \textbf{(C4)}.

\subsection{Expert Feedback}
\label{sec:feedback}
Our three domain experts were a team of graduate students broadly interested in computational work at the intersection of linguistics, learning sciences and NLP.
These experts were the same collaborators involved in the design process and carried out the deployment.
Some of the experts contributed to the development of the \textit{iTELL} framework, and all had experience working with \textit{iTELL}.
All studied or taught language and NLP in various ways, including work in linguistics, language learning, development, and teaching, and using LLMs to predict text readability, reading comprehension, personally identifiable information, and build other learning tools.
All three experts self-identified as male.
We organized the interview questions and domain experts' feedback around three broad questions we directly asked each collaborator.
We then conducted inductive thematic analysis \cite{Boyatzis:1998:ThematicAnalysis} of the feedback to identify emergent themes that were discussed amongst all authors.

\begin{enumerate}
    \item \textbf{Understanding LLMs} -- ``How did \textit{iScore} improve your understanding of LLM performance?''
    \item \textbf{Evaluating LLMs} -- ``How were the visualizations and interactions in \textit{iScore} useful for interpreting LLMs?''
    \item \textbf{Transparency and Trust in LLMs} -- ``How did \textit{iScore} enable transparency and promote trust in using LLMs?''
\end{enumerate}

\medskip
\noindent\textbf{(1) Understanding LLMs.  }
We first inquired about the major challenges the experts face in understanding LLMs.
We found several recurring themes: understanding how decisions are made from billions of LLM parameters; investigating whether LLMs use context around words; and ``seeing'' what is going on inside models using traditional programming tools.
There were also architecture-specific challenges related to Longformers, as one expert explained: \textit{``Since we are using Longformer with the sliding attention window, there are developer degrees of freedom that can be tweaked compared with other models.''}
These were important to address for several reasons.
One expert was concerned with the long-term reproducibility of their findings, feeling that \textit{``there is always anxiety or tension around whether an outcome is just a fluke.''}
Another described how addressing these challenges can also lead to new research questions: \textit{``Sometimes we couldn't explain [findings] with the visualizations, but it led us to ask new questions about our research, and it served to remind us that we don’t always know how [the LLMs] are working.''}

\textit{iScore} enabled the experts to learn several new things about their models.
Using the \textit{Token Attention} visualization, one expert discovered that punctuation tended to contain a lot of information in final model layers, as the marks were always underlined in dark purple.
Consequently, removing punctuation would cause scores for both Content and Wording models to go down drastically.
They were concerned with the ramifications of this finding: \textit{``We didn't expect to see attention weights on punctuation... The question is, does this have any pedagogical impact?''}
Another expert found that removing the first sentence from summaries would drop scores by nearly $90\%$.
They hypothesized that the models were focusing on the beginning of the sentence and the punctuation denoting the end of the first sentence, as the classification token is at the start of the input.
This may corroborate the findings from our case study, as the retrained models with global attention on the entire summary did not exhibit this behavior as strongly.
The expert reflected on this behavior, saying \textit{``While [topic sentences] could be important for writing summaries, it may not be the way the model is operating.''}

\medskip
\noindent\textbf{(2) Evaluating LLMs.  }
To discover these insights, the experts felt the \textit{Input Perturbation} and \textit{Token Attention} visualizations allowed them to see how things worked at a higher level.
One expert summarized how this helped them, saying \textit{``we now have methods to help us show that the models work as we expect, and also when we don't expect results.''}
For example, the same expert described how they used the \textit{Assignments Panel} and \textit{Input Perturbation} visualizations to create summary variations and analyze their models: \textit{``I was using input perturbations to illustrate how adversarial attacks can trick models. This led me to new research questions. The perturbations are one word at a time. What if we do n-gram changes?''}
Yet there was room for improvement, as another expert sometimes had trouble knowing what to look at in the \textit{Token Attention} visualization: \textit{``There seems to be a consistent noise in the attention. What is significant over and above the noise?''}
In the future, \textit{iScore} could include statistical analysis such as highlighting attentions far outside of the mean.

As they explained their process for discovering insights with \textit{iScore}, two of the experts developed new ideas for how the visualizations could help them close the loop of model development.
One used the \textit{Model Analysis View} to quickly profile and decide between the best-performing models: \textit{``I often train multiple models for a task and want to compare them. What I like to do is switch between the models using the drop-downs.''}
To promote reproducibility in the future and communicate their findings directly in the \textit{iScore} interface, they requested hypothesis testing capabilities such as a chi-squared test.
The other expressed a desire to use the \textit{Scores Dashboard} for tracking changes across a specific task, such as key phrase analysis: \textit{``It may be too easy to just throw in the key phrases and increase model scores and we don’t want that.''}
This type of analysis could generate additional examples for injecting into the LLM training data, to make them robust to learners trying to ``game'' the system by spamming the input with key phrases.

\medskip
\noindent\textbf{(3) Transparency and Trust in LLMs.  }
The experts all reported different factors and requirements for building trust.
Reproducibility was a requirement for one learning engineer, as they wanted to avoid unexpected surprises especially when deploying their LLMs: \textit{``You might get great results, and sometimes they look too good and you have to go back and check your code.''}
Addressing biases primarily in the training data was mentioned by another, though the expert also noted the difficult endeavor of explaining the vast number of LLM parameters responsible for producing a score as a barrier to realizing transparency.
Another expert believed in clearly demonstrating model limitations and edge cases, especially given the ability for LLMs to hallucinate: \textit{``If we are dead set on fooling models, we can create examples to fool the models.''}
Finally, the context of deployment was considered important for all experts, as one expert explained: \textit{``Convincing stakeholders becomes easier if you can show them examples of how the model is working.''}

The experts unanimously agreed that the most useful feature of \textit{iScore} for inspiring trust was the \textit{Input Perturbation} visualization.
One expert felt that, \textit{``in a way, input perturbation allows you to put in adversarial examples to test your trust in the model.''}
At the same time, another expert commented that the \textit{Token Attention} visualization did not achieve complete model transparency for them.
Following up, they still expressed optimism that \textit{``trust can be achieved even without full transparency''}, especially when using the \textit{Assignments Panel} and \textit{Input Perturbation} visualization: \textit{``The aspects of iScore that can test variations allow us to demonstrate what happens with changes in summaries. This can broadly allow us to improve trust in these systems.''}
This benefit extended outside of the development environment, allowing the experts to show their work in context of deployment with stakeholders.
On the opposite end of the spectrum, for developers deep in the NLP loop, one expert felt \textit{iScore} provided a layer between development and production where few tools exist to provide qualitative feedback to learning engineers at the intersection of educational data and machine learning.
For them, \textit{iScore} promoted trust through reproducibility because the parameters of each LLM can be quickly iterated, tested, and reported using screenshots of the system.

\section{Discussion}
\label{sec:discussion}

\subsection{Implications For Design}
\label{sec:design_implications}
By distilling the findings from our evaluation, we synthesized design implications for future visual analytics systems situated in terms of our key design challenges \textbf{(C)} outlined in Sect.~\ref{sec:design_challenges} as well as prior literature.

\medskip
\noindent\textbf{Structure LLM evaluation using visual hierarchies.}
We observed that communicating the LLM evaluation process through visual hierarchies enabled thoughtful comparison between LLM inputs and outputs, improved understanding of how models work and increased trust that models are working correctly.
To scaffold comparison, users should be able to easily upload and/or write multiple text inputs to their LLMs at once \textbf{(C1)}.
Text inputs should be arranged to visually communicate comparison (e.g., grouping by topic, version, author, etc.) and include options for automatically testing multiple models simultaneously \textbf{(C1)}, saving users time and aiding their sensemaking process.
Further, we found that visualizing the distribution of LLM outputs across groups of models, inputs and revisions helped users communicate valuable insights when reporting their findings.
One way to enrich visual comparison is visualizing distributions in context of the "ground truth" model training data \textbf{(C1)}.
Another way is to give users options for grouping and visualizing multiple types of automatic text revisions \textbf{(C2)}.
Finally, our users suggested visually highlighting and grouping potential relationships between all tokens using statistical tests \textbf{(C3, C4)}.
We discuss future work in this area in Sect.~\ref{sec:limitations_future_work}.

\medskip
\noindent\textbf{Scale LLM interpretability methods to large inputs.}
Supporting analysis of multiple large text inputs became both a computational and mental bottleneck, highlighting the challenge of scaling interpretability methods as LLMs are developed that can process larger inputs \cite{Xiao:2023:StreamingLLM}.
Leveraging fluid interaction \cite{Elmqvist:2011:FluidInteraction} can help users identify and track sets of tokens that a model is focusing on and compare these sets against other tokens at multiple levels of aggregation.
For example, users should be able to visualize and compare differences between text inputs at multiple levels of aggregation (e.g., token, word, sentence) as well as interactively switch between levels on the fly \textbf{(C2, C3)}.
Interactively specifying encodings for how changes to text spans affect LLM outputs using inline annotations (e.g., underlining tokens, words, sentences, etc.) allows users to read text naturally at scale while fluidly identifying patterns between individual and groups of annotations, even as inputs grow.
We found this to be useful both when externally probing models such as input perturbation and when analyzing internal model weights \textbf{(C3)}.
As interpretability methods can be computationally expensive for large numbers of tokens, users should also be able to subset and apply methods both simultaneously and independently to any number of inputs \textbf{(C1)}.
Result sets should then be easy to switch between on the fly, allowing users to quickly visualize and compare differences without slowing down the user experience \textbf{(C1, C3)}.

\medskip
\noindent\textbf{Potential future applications.}
Taken together, these design implications inspire future applications of visual analytics for interpreting LLMs.
For example, consider an engineer training LLMs for a question-answer module, or a NLP researcher investigating LLMs trained to assign a readability score.
They could import their models into \textit{iScore} to interpret what parts of the text the models are paying attention to using the \textit{Token Attention} visualization, as well as how the models respond to different perturbations of the input text using the \textit{Input Perturbation} visualization.
Developers could also deploy the perturbation features with users to collect feedback for retraining their models, by analyzing how inputs are revised and how scores change as users respond to seeing model scores.

\subsection{Responsible and Ethical AI for Education}
\label{sec:pedagogy_responsible_ai}
The scalability and generalizability of LLMs could enable summary writing assignments to be used in a wider variety of environments, including Massive Open Online Courses (MOOCs), learning platforms and intelligent textbooks \cite{gamage2021peer}.
To realize this, there are several broad practical and ethical considerations of using machine learning (ML) and artificial intelligence (AI) that \textit{iScore} addresses, helping learning engineers deploy responsible AI in educational tools like \textit{iTELL} and mitigate the pedagogical risks of using AI-powered tools in critical learning environments \cite{kasneci2023chatgpt}.

It is necessary to remove potential biases and evaluate the fairness of any AI model's predictions.
Ensuring fairness means using balanced data sets during the training period as well as oversight and regular re-training based on the performance of AI models in production.
Using \textit{iScore}, learning engineers can rapidly explore and identify potential biases that manifest either internally in the model weights or externally in the model scores.
For example, they could track patterns of biases in score assignments across revisions containing different personal identity phrases such as pronouns.
AI-powered tools for education should also accommodate the target language of learners by providing both equitable and interpretable outputs.
This can be achieved by expanding LLM training across a variety of data sets in various languages, and using the input and history tracking features of \textit{iScore} to evaluate LLM performance at scale across many different summaries.
Finally, perceptions of AI are an important factor for educators in the adoption of AI-powered education technologies, with trust being one of the most important determinants of whether an AI model is actually used \cite{choi2023influence}.
One of the greatest barriers to building trust is a lack of technical expertise among educators in understanding what AI is capable of.
We believe \textit{iScore} contributes practical methods for enabling transparency and helping non-ML-experts build trust in deploying AI models like LLMs in tools like \textit{iTELL}.
For example, the perturbation features of \textit{iScore} could be adapted and simplified for presentation to non-ML-expert audiences, such as instructors and learners, as a hands-on version that seeks to promote trust in using AI-powered educational technology.
In the fields of education, machine learning and HCI broadly, it is imperative to develop tools like \textit{iScore} that enable engineers to build ethical and responsible AI tools and interpret them well enough to begin building the essential trust necessary to see them implemented safely in learning environments.

\subsection{LLM Generalizability}
\label{sec:generalizability}
In this paper, we demonstrated the capabilities of \textit{iScore} using Longformer, a pre-trained LLM based on RoBERTa, as Longformer allows for longer text inputs that combine summary writing and source text during inference.
RoBERTa itself is a fundamental language model that continues to obtain state-of-the-art results on classification and natural language understanding tasks \cite{Srivastava:2022:BeyondImitationGame}.
Beyond RoBERTa, \textit{iScore} can be used to interpret any transformer-based model that utilizes a separator token for sequence classification, such as a reference text and a summary or a question and an answer, building on the successes of other transformer-based LLM architectures.
This includes both encoder- and decoder-only models as well as encoder-decoder models such as GPT-3 \cite{Brown:2020:GPT3} and T5 \cite{Raffel:2020:T5}.
To enable larger input text sizes, StreamingLLM \cite{Xiao:2023:StreamingLLM} is a recent framework that can replicate the larger input text size capabilities of Longformers using attention sinks for models such as LLaMa \cite{Touvron:2023:Llama}.
Finally, to make our methods accessible to learning engineers and NLP researchers in the future, we use transformers downloaded from the HuggingFace Transformers \cite{Wolf:2020:HuggingfaceTransformers} model library that are open-source and lightweight enough to run on a single GPU.

\subsection{Limitations and Future Work}
\label{sec:limitations_future_work}
Several analytical limitations were expressed during the interviews.
While our collaborators were positive about moving towards using visualizations that show model weights, they still had issues with interpreting them.
One felt that there was a constant noise in the perturbation scores and attention weights that made it difficult to identify important patterns and relationships.
Interactions at specific layers/heads were also difficult to explain, as one engineer disliked scanning across layers/heads to perform an ``axiomatic'' type of analysis.
While attention visualizations enabled our collaborators to improve model understanding and accuracy in Sect.~\ref{sec:case_study}, there is on-going debate whether attention weights in transformers can be used as a reliable source of interpretation for model performance \cite{Clark:2019:AnalyzingBERTAttention, Jain:2019:AttentionIsNotExplanation, Wiegreffe:2019:AttentionIsNotNotExplanation, Atanasova:2020:ExplainingUsingWeights}.
Several engineers suggested augmenting \textit{iScore} with statistical tests that could improve how to interpret and compare attentions and perturbations.
These could include hypothesis tests such as a chi-squared test to compare distributions of attention weights between models or a score outlier detection method using interquartile ranges.
Further, our collaborators found the input perturbation methods to be the most useful, yet this was also the most computationally demanding method, which can limit the responsiveness and increase deployment requirements for this feature.
One promising direction to emulate perturbation is performing gradient-based attribution of input features to the final model prediction \cite{Sundararajan:2017:IntegratedGradients}, by developing new methods specifically for the sliding attention window in Longformer models.
Perturbation also only considers the marginal contribution of a token/sentence among the remaining ones, ignoring interactions among features (i.e., token/sentence). 
Using Shapley Values (e.g., SHAP \cite{Lundberg:2017:SHAP}) may be a more accurate approach, albeit at the cost of increased computational time over the already expensive perturbations.
Finally, one collaborator requested a method for performing automatic sentence-level replacement, to investigate how this could improve writing feedback in the context of automatic scoring.
This could be achieved by using additional LLMs to generate grammatically and/or semantically similar sentences for replacement.

\section{Conclusion}
\label{sec:conclusion}

As educational technologies increasingly utilize ``black box'' deep learning models, it is critical to empower model developers to understand, evaluate, and build trust in their models before deploying them in critical learning environments.
In this work, we conducted a user-centered design process with a team of learning engineers who train and deploy large language models (LLMs) that automatically score written summaries of source texts.
We presented \textit{iScore}, a visual analytics system for learning engineers to upload, compare and visualize multiple LLM-scored summaries.
By structuring evaluation of LLM performance around comparison of multiple source/summary pairs, visualizing the provenance of summary scores in context of the LLM training data, and presenting different LLM interpretability methods simultaneously at multiple scales, \textit{iScore} increases LLM transparency and trust throughout human-in-the-loop evaluation of LLM performance.

\begin{acks}
  This material is based upon work supported by the National Science Foundation under Grant No. 2247790 and Grant No. 2112532.
  Any opinions, findings, and conclusions or recommendations expressed in this material are those of the author(s) and do not necessarily reflect the views of the National Science Foundation.
\end{acks}

\bibliographystyle{ACM-Reference-Format}
\bibliography{main}

\end{document}